\documentclass[pra,aps,amsmath,amssymb,showkeys]{revtex4}
\usepackage{graphicx}
\newcommand{\hh}{{\mathcal{H}}}
\newcommand{\tr}{\mathrm{tr}}
\newcommand{\rmA}{{\mathrm{A}}}
\newcommand{\rmB}{{\mathrm{B}}}
\newcommand{\rmH}{{\mathrm{H}}}
\newcommand{\rmR}{{\mathrm{R}}}
\newcommand{\pen}{\openone}
\newcommand{\bro}{\boldsymbol{\rho}}
\newcommand{\me}{{\mathsf{E}}}
\newcommand{\mn}{{\mathsf{M}}}
\newcommand{\nm}{{\mathsf{N}}}
\newcommand{\xdif}{{\mathrm{d}}}
\newcommand{\cald}{{\mathcal{D}}}
\newcommand{\cle}{{\mathcal{E}}}
\newcommand{\clm}{{\mathcal{M}}}
\newcommand{\cln}{{\mathcal{N}}}
\newcommand{\calp}{{\mathcal{P}}}
\newcommand{\dn}{{\mathbb{D}}}
\newcommand{\psyms}{{\mathsf{\Pi}}_{\mathrm{sym}}^{(s)}}
\newcommand{\psymt}{{\mathsf{\Pi}}_{\mathrm{sym}}^{(t)}}

\newcommand{\veps}{\varepsilon}
\newcommand{\wta}{{\tilde{a}}}
\newcommand{\wb}{{\tilde{b}}}
\newcommand{\vcp}{{\mathtt{p}}}
\newcommand{\vark}{\varkappa}

\begin{document}
\clearpage
\preprint{}

\title{Complementarity relations for design-structured POVMs in terms of generalized entropies of order $\alpha\in(0,2)$}

\author{Alexey E. Rastegin}
\email{alexrastegin@mail.ru}
\affiliation{Department of Theoretical Physics, Irkutsk State University, K. Marx St. 1, Irkutsk 664003, Russia}

\begin{abstract}
Information entropies give a genuine way to characterize
quantitatively an incompatibility in quantum measurements. Together
with the Shannon entropy, few families of parametrized entropies
have found use in various questions. It is also known that a
possibility to vary the parameter can often provide more
restrictions on elements of probability distributions. In quantum
information processing, one often deals with measurements having
some special structure. Quantum designs are currently the subject of
active research, whence the aim to formulate complementarity
relations for related measurements occurs. Using generalized
entropies of order $\alpha\in(0,2)$, we obtain uncertainty and
certainty relations for POVMs assigned to a quantum design. The
structure of quantum designs leads to several restrictions on
generated probabilities. We show how to convert these restrictions
into two-sided entropic estimates. One of the used ways is based on
truncated expansions of the Taylor type. The recently found method
to get two-sided entropic estimates uses polynomials with flexible
coefficients. We illustrate the utility of this method with respect
to both the R\'{e}nyi and Tsallis entropies. Possible applications
of the derived complementarity relations are briefly discussed.
\end{abstract}


\maketitle

\pagenumbering{arabic}
\setcounter{page}{1}

\section{Introduction}\label{sec1}

The Heisenberg uncertainty principle \cite{heisenberg} is now
interpreted as a scientific concept of general applicability.
Quantum-mechanical complementarity relations are interesting in many respects.
They play a certain role in building feasible schemes for some tasks
of quantum information, for instance, for entanglement detection,
quantum tomography and steering. There exist several ways to
characterize quantitatively potential uncertainties in various
quantum measurements. The traditional approach to preparation
uncertainty relations deals with the product of variances
\cite{kennard,robert}. It has later been criticized for several
reasons \cite{deutsch,maass}. Entropic relations have been proposed
as a proper way to characterize quantum uncertainties. For
finite-dimensional observables, basic results in this area are
resumed in \cite{ww10,cbtw17}. Entropic uncertainty relations for
continuous variables are reviewed in \cite{brud11,cerf19}.

Special types of measurements are widely used in quantum information
science. Mutually unbiased bases \cite{bz10,mnw24} and symmetric
informationally complete measurements \cite{rbksc04} are especially
important examples. They are linked to rather difficult questions
about discrete structures in Hilbert space (see, e.g., the reviews
\cite{fhs2017,fivo20} and references therein). Quantum and unitary
designs are considered as a useful tool in quantum information
processing
\cite{ambain07,gross07,scottjpa8,dcel9,roy09,cirac18,bhmn19,kwg19,chart19}.
Formally, quantum design is a complex projective $t$-design
\cite{scottjpa}. To the given quantum designs, we can assign one or
several measurements with many restriction on generated
probabilities. For such measurements, uncertainty relations in terms
of $\alpha$-entropies of order $\alpha\geq2$ were obtained in
\cite{guhne20,rastdes}. The question of deriving Shannon-entropy
complementarity relations for design-structured POVMs was recently
examined in \cite{rastpol} and applied to quantum coherence in
\cite{rastunc}.

For integer values $\alpha\geq2$, uncertainty relations directly
follow from properties of quantum designs. The paper \cite{rastpol}
uses truncated expansions of the Taylor type and polynomials with
flexible coefficients. The latter has been motivated by Lanczos
\cite{lanczos} as a good method to get oscillating approximations.
As was shown in \cite{rastpol}, polynomials with flexible
coefficients also lead to fitting the function of interest from
below or above solely. The aim of this work is to study the
question, whether this method is suitable for generalized entropies.
The paper is organized as follows. In Sect. \ref{sec2}, the
required material on quantum designs is given. In Sect.
\ref{sec3}, we derive two-sided estimates on a power function
naturally connected with the considered entropies. In Sect.
\ref{sec4}, complementarity relations for design-structured POVMs in
terms of generalized entropies are formulated. Several applications
of the derived relations are briefly discussed in Sect.
\ref{sec5}. In Sect. \ref{sec6}, we conclude the paper. Appendix
\ref{aniq2} summarizes auxiliary material.

\section{Preliminaries}\label{sec2}

In this section, we recall some facts about quantum designs. In
$d$-dimensional Hilbert space $\hh$, we consider lines passing
through the origin. These lines form a complex projective space
\cite{scottjpa}. Up to a phase factor, each line is represented by a
unit vector $|\phi\rangle\in\hh$. The set
$\dn=\bigl\{|\phi_{k}\rangle:\,\langle\phi_{k}|\phi_{k}\rangle=1,\,k=1,\ldots,K\bigr\}$
is a quantum $t$-design, when for all real polynomials $\calp_{t}$
of degree at most $t$ it holds that
\begin{equation}
\frac{1}{K^{2}}\sum_{j,k=1}^{K}
\calp_{t}\Bigl(\bigl|\langle\phi_{j}|\phi_{k}\rangle\bigr|^{2}\Bigr)=
\int\!\int\xdif\mu(\psi)\,\xdif\mu(\psi^{\prime})\>
\calp_{t}\Bigl(\bigl|\langle\psi|\psi^{\prime}\rangle\bigr|^{2}\Bigr)
\, . \label{tdesdf}
\end{equation}
Here, $\mu(\psi)$ denotes the unique unitarily invariant probability
measure on the corresponding complex projective space
\cite{scottjpa}. Obviously, each $t$-design is also a $s$-design
with $s\leq{t}$. It was proved in \cite{seym1984} that $t$-designs
in a path-connected topological space exist for all $t$ and $d$. On
the other hand, there is no common strategy to generate designs in
all required cases. Nevertheless, there exist many interesting
examples used in applications \cite{hardin96}. Quantum designs are
linked to the problem of building SIC-POVMs and tight rank-one
informationally complete POVMs \cite{rbksc04,scottjpa}.

The following properties of quantum designs will be used. Let $\psymt$ be
the projector onto the symmetric subspace of $\hh^{\otimes{t}}$. The
trace of $\psymt$ gives dimensionality of this symmetric subspace. It
holds that \cite{scottjpa}
\begin{equation}
\frac{1}{K}\>\sum_{k=1}^{K}
|\phi_{k}\rangle\langle\phi_{k}|^{\otimes{t}}=\cald_{d}^{(t)}\,\psymt
\, , \label{topys}
\end{equation}
where $\cald_{d}^{(t)}$ denotes the inverse of
$\tr\bigl(\psymt\bigr)$, namely
\begin{equation}
\cald_{d}^{(t)}=\binom{d+t-1}{t}^{\!-1}
=\frac{t!\,(d-1)!}{(d+t-1)!}
\ . \label{indim}
\end{equation}
At the given $t$, we can rewrite (\ref{topys}) for all positive
integers $s\leq{t}$. Substituting $t=1$ leads to the formula
\begin{equation}
\frac{d}{K}\>\sum_{k=1}^{K}
|\phi_{k}\rangle\langle\phi_{k}|=\pen_{d}
\, . \label{comprel}
\end{equation}
Thus, unit vectors $|\phi_{k}\rangle$ allow us to build to a
resolution of the identity in $\hh$. In principle, there may be
several resolutions assigned to the given $t$-design. We cannot list
all of them {\it a priori}, without an explicit analysis of the
design. Obviously, one can take the complete set $\cle$ consisting
of operators
\begin{equation}
\me_{k}=\frac{d}{K}\>
|\phi_{k}\rangle\langle\phi_{k}|
\, . \label{mkdef}
\end{equation}
Sometimes, $M$ rank-one POVMs $\bigl\{\cle^{(m)}\bigr\}_{m=1}^{M}$
can be assigned to the given quantum design. Each of them
consist of $\ell$ operators of the form
\begin{equation}
\me_{j}^{(m)}=\frac{d}{\ell}\>
|\phi_{j}^{(m)}\rangle\langle\phi_{j}^{(m)}|
\, . \label{mejdf}
\end{equation}
The integers $\ell$ and $M$ are connected by $K=\ell{M}$.

When the pre-measurement state is described by density matrix
$\bro$, the probability of $j$-th outcome is equal to
\begin{equation}
p_{j}(\cle^{(m)};\bro)=\frac{d}{\ell}\,\langle\phi_{j}^{(m)}|\bro|\phi_{j}^{(m)}\rangle
\, . \label{prbk}
\end{equation}
Due to (\ref{topys}), for all $s=2,\ldots,t$ one has \cite{guhne20}
\begin{equation}
\frac{1}{K}\>\sum_{k=1}^{K}
\langle\phi_{k}|\bro|\phi_{k}\rangle^{s}=\cald_{d}^{(s)}\tr\bigl(\bro^{\otimes{s}}\psyms\bigr)
\, . \label{indet}
\end{equation}
Combining (\ref{prbk}) with (\ref{indet}) then gives
\begin{equation}
\sum_{m=1}^{M}\sum_{j=1}^{\ell}p_{j}(\cle^{(m)};\bro)^{s}=
\left(\frac{d}{\ell}\right)^{\!s}\,\sum_{k=1}^{K}\langle\phi_{k}|\bro|\phi_{k}\rangle^{s}=
K\ell^{-s}d^{\,s}\,\cald_{d}^{(s)}\tr\bigl(\bro^{\otimes{s}}\psyms\bigr)
\, . \label{mindt}
\end{equation}
When a single POVM is assigned, one has $\ell=K$ and
\begin{equation}
\sum_{k=1}^{K}p_{k}(\cle;\bro)^{s}=
K^{1-s}d^{\,s}\,\cald_{d}^{(s)}\tr\bigl(\bro^{\otimes{s}}\psyms\bigr)
\, . \label{indek}
\end{equation}
The authors of \cite{guhne20,cirac18} have answered the question how
to write $\tr\bigl(\bro^{\otimes{s}}\psyms\bigr)$ as a sum of
monomials of the moments of $\bro$. Complexity of such
expressions increases with growth of $s$. To avoid bulky expressions
in the following, one introduces the quantities
\begin{align}
\bar{\beta}_{\ell}^{(s)}(\bro)&=\ell^{1-s}d^{\,s}\,\cald_{d}^{(s)}\tr\bigl(\bro^{\otimes{s}}\psyms\bigr)
\, , \label{betn}\\
\bar{\beta}^{(s)}(\bro)&=K^{1-s}d^{\,s}\,\cald_{d}^{(s)}\tr\bigl(\bro^{\otimes{s}}\psyms\bigr)
\, . \label{betk}
\end{align}
These terms will also be used for $s=0,1$, when
$\bar{\beta}^{(0)}(\bro)=K$ and $\bar{\beta}^{(1)}(\bro)=1$. Due to
(\ref{mindt}) and (\ref{indek}), restrictions on measurement
statistics for design-structured POVMs are imposed. Hence, entropic
uncertainty relations for such measurement follow. Using R\'{e}nyi
entropies of order $\alpha\geq2$, uncertainty relations were
obtained in \cite{guhne20,rastdes}. The paper \cite{rastdes} deals
with relations improved by further developing the idea of
\cite{rastmubs}. The authors of \cite{guhne20} also formulated
uncertainty relations in terms of Tsallis entropies. Since the
R\'{e}nyi entropy cannot increase with growth of its order, the
results of \cite{guhne20,rastdes} also give lower bounds on the
corresponding Shannon entropy. However, such estimates are
sufficiently weak. The problem of deriving uncertainty relations in
terms of the Shannon entropy lies beyond the methods of
\cite{guhne20,rastdes}. In effect, this question has been examined
in the paper \cite{rastpol}. We shall now extend the methods of
\cite{rastpol} to generalized entropies of order between $0$ and
$2$.

\section{On two-sided estimating entropic functions}\label{sec3}

In this section, we will derive two-sided estimates on the entropic
functions of interest in terms of the power sums of probabilities.
Generalized $\alpha$-entropies can be defined in terms of the
function
\begin{equation}
\eta_{\alpha}(x):=\frac{x^{\alpha}-x}{1-\alpha}=-x^{\alpha}\ln_{\alpha}(x)
\ , \label{yalp}
\end{equation}
where $\alpha>0$. The $\alpha$-logarithm of positive $x$ is commonly defined as
\begin{equation}
\ln_{\alpha}(x):=
\begin{cases}
 \frac{x^{1-\alpha}-1}{1-\alpha}\>, & \text{ for } 0<\alpha\neq1 \, , \\
 \ln{x}\, , & \text{ for } \alpha=1 \,.
\end{cases}
\label{lanl}
\end{equation}
We aim to estimate (\ref{yalp}) from two sides for $\alpha\in(0,2)$
and $x\in[0,1]$. We first suppose that $\beta=1-\alpha\in(0,1)$. By
the Taylor expansion, one gets
\begin{equation}
(1-z)^{-\beta}=1+\sum_{r=1}^{\infty}\frac{\beta^{\bar{r}}z^{r}}{r!}
\ . \label{risf}
\end{equation}
where $\beta^{\bar{r}}=\beta(\beta+1)\cdots(\beta+r-1)$ is the
rising factorial \cite{GKP94}. Substituting $z=1-x$ into
(\ref{risf}), we then write
\begin{equation}
\eta_{\alpha}(x)=\frac{x(x^{-\beta}-1)}{\beta}=
\frac{x}{1-\alpha}\,\sum_{r=1}^{\infty}\frac{(1-\alpha)^{\bar{r}}}{r!}\,(1-x)^{r}
\, , \label{yaab}
\end{equation}
where $n\geq2$. All the summands in the right-hand side of
(\ref{yaab}) are positive, whence
\begin{equation}
\frac{x^{\alpha}-x}{1-\alpha}\geq\frac{x}{1-\alpha}\,\sum_{r=1}^{n-1}\frac{(1-\alpha)^{\bar{r}}}{r!}\,(1-x)^{r}
=\sum_{s=1}^{n}a_{n\alpha}^{(s)}\,x^{s}
\, . \label{yaab1}
\end{equation}
To check this conclusion for $\alpha\in(1,2)$, we take
$\gamma=\alpha-1\in(0,1)$. Using the Taylor expansion
\begin{equation}
(1-z)^{\gamma}=1-\gamma{z}\sum_{r=0}^{\infty}\frac{(1-\gamma)^{\bar{r}}z^{r}}{(r+1)!}
\ , \label{risf1}
\end{equation}
one has
\begin{equation}
\eta_{\alpha}(x)=\frac{x(1-x^{\gamma})}{\gamma}=
x\sum_{r=0}^{\infty}\frac{(2-\alpha)^{\bar{r}}}{(r+1)!}\,(1-x)^{r+1}
\, . \label{yagb}
\end{equation}
Formally, the latter sum coincides with (\ref{yaab}), and its 
summands are all positive as well. Therefore, the inequality
(\ref{yaab1}) holds for $\alpha\in(0,1)\cup(1,2)$ and $n\geq2$. This
inequality will be used to estimate $\alpha$-entropies from below.

Let us obtain inequalities for estimating $\alpha$-entropies from
above. For $\alpha\in(0,1)$, we rewrite (\ref{risf1}) with $\alpha$
instead of $\gamma$, whence
\begin{equation}
x^{\alpha}-x
=1-x-\alpha(1-x)\sum_{r=0}^{\infty}\frac{(1-\alpha)^{\bar{r}}}{(r+1)!}\,(1-x)^{r}
\leq1-x-\alpha(1-x)\sum_{r=0}^{n-1}\frac{(1-\alpha)^{\bar{r}}}{(r+1)!}\,(1-x)^{r}
\, . \label{ybab}
\end{equation}
To estimate $\alpha$-entropies from above for $\alpha\in(0,1)$, we
will use the inequality
\begin{equation}
\frac{x^{\alpha}-x}{1-\alpha}\leq\frac{1-x}{1-\alpha}
\,\biggl(
1-\alpha\sum_{r=0}^{n-1}\frac{(1-\alpha)^{\bar{r}}}{(r+1)!}\,(1-x)^{r}
\biggr)=\sum_{s=0}^{n}b_{n\alpha}^{(s)}\,x^{s}
\, . \label{ybab1}
\end{equation}
The inequality (\ref{ybab1}) remains valid for $\alpha\in(1,2)$.
Here, we write the expansion
\begin{equation}
(1-z)^{\alpha}=1-\alpha{z}+\alpha\gamma{z}^{2}\sum_{r=0}^{\infty}\frac{(2-\alpha)^{\bar{r}}z^{r}}{(r+2)!}
\ , \label{palpy}
\end{equation}
so that
\begin{equation}
x-x^{\alpha}
=x-1+\alpha(1-x)-\alpha\gamma\sum_{r=0}^{\infty}\frac{(2-\alpha)^{\bar{r}}(1-x)^{r+2}}{(r+2)!}
\leq{x}-1+\alpha(1-x)-\alpha\gamma\sum_{r=0}^{n-2}\frac{(2-\alpha)^{\bar{r}}(1-x)^{r+2}}{(r+2)!}
\ . \label{baby}
\end{equation}
Dividing the latter by $\gamma=\alpha-1$ gives the right-hand side
of (\ref{ybab1}). Thus, the inequality (\ref{ybab1}) holds
$\alpha\in(0,1)\cup(1,2)$ and $n\geq2$.

Doing some calculations, the explicit expressions for the
coefficients reads as
\begin{align}
& a_{n\alpha}^{(1)}=\frac{1}{1-\alpha}\>\sum_{r=1}^{n-1}\frac{(1-\alpha)^{\bar{r}}}{r!}
\ , \qquad
& a_{n\alpha}^{(s)}=\frac{(-1)^{s-1}}{1-\alpha}\sum_{r=s-1}^{n-1}
\frac{(1-\alpha)^{\bar{r}}}{r!}\>\binom{r}{s-1}
\qquad
(2\leq{s}\leq{n})
\, , \label{aceff}\\
& b_{n\alpha}^{(0)}=1-\frac{\alpha}{1-\alpha}\sum_{r=1}^{n-1}\frac{(1-\alpha)^{\bar{r}}}{(r+1)!}
\ , \qquad
& b_{n\alpha}^{(1)}=\alpha\,a_{n\alpha}^{(1)}-1
\ , \qquad
b_{n\alpha}^{(s)}=\frac{\alpha}{s}\>a_{n\alpha}^{(s)}
\qquad
(2\leq{s}\leq{n})
\, . \label{bceff}
\end{align}
In the limit $\alpha\to1$, the above expressions reduce to the
coefficients derived in \cite{rastpol}. Namely, one has
$b_{n1}^{(0)}=1/n$ and
\begin{align}
&a_{n1}^{(1)}=\sum_{r=1}^{n-1}\frac{1}{r}
\ , \qquad
&a_{n1}^{(s)}=(-1)^{s-1}\sum_{r=s-1}^{n-1}\frac{1}{r}\>\binom{r}{s-1}
\qquad
(2\leq{s}\leq{n})
\, , \label{aeff}\\
&b_{n1}^{(1)}=\sum_{r=2}^{n-1}\frac{1}{r}
\, , \qquad
&b_{n1}^{(s)}=\frac{(-1)^{s-1}}{s}\sum_{r=s-1}^{n-1}\frac{1}{r}\>\binom{r}{s-1}
\qquad
(2\leq{s}\leq{n})
\, . \label{boeff}
\end{align}
It was shown in \cite{rastpol} that, for $n\geq2$, we have the
two-sided estimate
\begin{equation}
\sum_{s=1}^{n}a_{n1}^{(s)}\,x^{s}\leq-\,x\ln{x}\leq\frac{1}{n}
+\sum_{s=1}^{n}b_{n1}^{(s)}\,x^{s}
\, . \label{alp1}
\end{equation}

It is instructive to consider the limit $\alpha\to+0$, when the term
$(x^{\alpha}-x)/(1-\alpha)$ reduces to $1-x$. We directly see from
(\ref{bceff}) that $b_{n0}^{(0)}=1$, $b_{n0}^{(1)}=-1$ and
$b_{n0}^{(s)}=0$ for $2\leq{s}\leq{n}$. In other words, the
inequality (\ref{ybab1}) is saturated in this limit. Further, the
coefficients (\ref{aceff}) are not expressed so simply. For
$\alpha\to+0$, the inequality (\ref{yaab1}) is transformed into the
form
\begin{equation}
1-x\geq(1-x)\bigl[1-(1-x)^{n-1}\bigr]
\, . \label{al0low}
\end{equation}
The left-hand side of (\ref{al0low}) gives a good estimate in some
left vicinity of the point $x=1$, but becomes insufficient for small
$x$. Explicit formulas for the coefficients $a_{n0}^{(s)}$ are
easily derived right from (\ref{al0low}). We refrain from presenting
the details here.

For $\alpha=2$ the
function of interest reads as $x-x^{2}$. Here, the right-hand side of
(\ref{yaab1}) reduces just to the same function, as the rising
factorial $(-1)^{\bar{r}}$ is equal to $(-1)$ for $r=1$ and zero for
all $r\geq2$. In the formula (\ref{ybab1}), the sum with respect to
$r$ includes nonzero terms only for $r=0,1$, whence we again have
$x-x^{2}$. That is, for $\alpha=2$ our two-sided estimate
is saturated in both its sides.

As was shown in \cite{rastpol}, polynomial expansions with flexible
coefficients sometimes work better than truncated expansions due to
the Taylor scheme. For $n\geq2$ and $x\in[0,1]$, it holds that
\cite{rastpol}
\begin{equation}
x\ln{x}\leq\frac{(-1)^{n}}{2n^{2}}\,\sum_{s=2}^{n} c_{n}^{(s)}\,\frac{x^{s}-x}{s-1}
\ . \label{yleg}
\end{equation}
The $n$-th shifted Chebyshev polynomial reads as
$T_{n}^{*}(x)=T_{n}(2x-1)$, where $T_{n}(\xi)$ is the $n$-th
Chebyshev polynomial of the first kind. It is well known that
\cite{lanczos}
\begin{equation}
T_{n}^{*}(x)=\sum\nolimits_{s=0}^{n} c_{n}^{(s)}x^{s}
\, , \qquad
c_{n}^{(s)}=(-1)^{n+s}\,2^{2s-1}
\!\left[
2\binom{n+s}{n-s}-\binom{n+s-1}{n-s}
\right]
 . \label{coeff}
\end{equation}
In particular, the first two coefficients are $c_{n}^{(0)}=(-1)^{n}$
and $c_{n}^{(1)}=(-1)^{n+1}2n^{2}$. Keeping in mind the limit
\begin{equation}
-\,\eta_{\alpha}(x)=\frac{x-x^{\alpha}}{1-\alpha}\underset{\alpha\to1}{\longrightarrow}x\ln{x}
\, , \label{laax}
\end{equation}
we will generalize (\ref{yleg}) to the functions of interest. The
function $x\mapsto-\,x^{\alpha}$ satisfies the first-order
differential equation
\begin{equation}
xy^{\prime}(x)-\alpha{y}(x)=0
\, . \label{fode1}
\end{equation}
Following Lanczos \cite{lanczos}, we now modify the right-hand side
of (\ref{fode1}) by additive term $\propto{T}_{n}^{*}(x)$. For
integer $s\geq2$, one further solves the equation
\begin{equation}
xq_{s\alpha}^{\,\prime}(x)-\alpha{q}_{s\alpha}(x)
=x^{s}
 \label{fodes}
\end{equation}
and obtains
\begin{equation}
q_{s\alpha}(x)=\frac{x^{s}}{s-\alpha}
\ . \label{qspol}
\end{equation}
Using these functions, we write the following sum
\begin{equation}
Q_{n\alpha}(x)=\sum_{s=2}^{n}\frac{c_{n}^{(s)}x^{s}}{s-\alpha}
\ , \label{qnsun}
\end{equation}
for which
\begin{equation}
xQ_{n\alpha}^{\,\prime}(x)-\alpha{Q}_{n\alpha}(x)
=\sum_{s=2}^{n} c_{n}^{(s)}x^{s}=T_{n}^{*}(x)-c_{n}^{(0)}-c_{n}^{(1)}x
\, . \label{foden}
\end{equation}
The ansatz (\ref{qnsun}) will be used to approximate (\ref{yalp}) by
polynomials. Following \cite{rastpol}, we also observe that
$-\,\eta_{\alpha}(x)$ vanishes for $x=1$. Hence, we correct
(\ref{foden}) by an additive linear term and multiply the result by
factor $\tau_{n\alpha}$ yet unknown. This step leads to the
polynomial
\begin{equation}
F_{n\alpha}(x)=\tau_{n\alpha}\sum_{s=2}^{n} c_{n}^{(s)}\,\frac{x^{s}-x}{s-\alpha}
\ . \label{gnsun}
\end{equation}
Up to a total factor, the latter sum reduces to (\ref{yleg}) for
$\alpha=1$. It is directly checked that
\begin{equation}
xF_{n\alpha}^{\,\prime}(x)-\alpha{F}_{n\alpha}(x)=\tau_{n\alpha}\sum_{s=2}^{n} c_{n}^{(s)}x^{s}
-x\tau_{n\alpha}(1-\alpha)Q_{n\alpha}(1)
\ . \label{rfod}
\end{equation}
The final step is to choose the factor $\tau_{n\alpha}$
appropriately. Let us consider the difference
\begin{equation}
\varDelta_{n\alpha}(x)=F_{n\alpha}(x)-\frac{x-x^{\alpha}}{1-\alpha}
\ . \label{vdiff}
\end{equation}
Substituting the latter into the left-hand side of (\ref{fode1})
finally gives
\begin{equation}
x\varDelta_{n\alpha}^{\,\prime}(x)-\alpha\varDelta_{n\alpha}(x)=
\tau_{n\alpha}\bigl[\,T_{n}^{*}(x)-c_{n}^{(0)}\bigr]-x\bigl[c_{n}^{(1)}\tau_{n\alpha}+(1-\alpha)Q_{n\alpha}(1)\tau_{n\alpha}+1\bigr]
\, . \label{dediff}
\end{equation}
Let us demand vanishing the second term in the right-hand side of
(\ref{dediff}), whence
\begin{equation}
\frac{1}{\tau_{n\alpha}}=(-1)^{n}\,2n^{2}-(1-\alpha)\sum_{s=2}^{n}\frac{c_{n}^{(s)}}{s-\alpha}=
(\alpha-1)\sum_{s=1}^{n}\frac{c_{n}^{(s)}}{s-\alpha}
\ . \label{invtau}
\end{equation}
For $\alpha=1$, the latter gives $\tau_{n1}=(-1)^{n}/(2n^{2})$, as
it stands in the right-hand side of (\ref{yleg}). Dividing
(\ref{dediff}) by $x^{1+\alpha}$ leads to
\begin{equation}
\frac{\xdif}{\xdif{x}}\>\frac{\varDelta_{n\alpha}(x)}{x^{\alpha}}=
\frac{(-1)^{n}\,\tau_{n\alpha}}{x^{1+\alpha}}\,\bigl[(-1)^{n}\,T_{n}^{*}(x)-1\bigr]
\ . \label{moden}
\end{equation}
Substituting $x=\cos^{2}\theta/2$ with $\theta$ between $0$ and
$\pi$, one has $T_{n}^{*}(x)=T_{n}(\cos\theta)=\cos{n}\theta$ and
\begin{equation}
(-1)^{n}\,T_{n}^{*}(x)-1=(-1)^{n}\cos{n}\theta-1\leq0
\, . \label{moden1}
\end{equation}
For $0<x<1$, the right-hand side of (\ref{moden}) is either positive
or negative. That is, the derivative of
$x^{-\alpha}\varDelta_{n\alpha}(x)$ does not change its sign. Now,
we can prove that the difference (\ref{vdiff}) cannot be negative
for $\alpha\in(0,2)$.

\newtheorem{eta02}{Proposition}
\begin{eta02}\label{reseta}
Let polynomial $F_{n\alpha}(x)$ be defined by (\ref{gnsun}) for
integer $n\geq2$ and real $\alpha\in(0,2)$. For all $x\in[0,1]$, it
holds that
\begin{equation}
\frac{x-x^{\alpha}}{1-\alpha}\leq{F}_{n\alpha}(x)
\, . \label{ineq01}
\end{equation}
\end{eta02}

{\bf Proof.} The case $\alpha=1$ has been considered in
\cite{rastpol}. Let us begin with the interval $\alpha\in(0,1)$. For
sufficiently small $\veps>0$, we have
\begin{equation}
\left.\frac{\varDelta_{n\alpha}(x)}{x^{\alpha}}
\,\right|_{x=\veps}=
\frac{1}{1-\alpha}-\veps^{1-\alpha}\biggl(\tau_{n\alpha}\sum_{s=2}^{n}\frac{c_{n}^{(s)}}{s-\alpha}+\frac{1}{1-\alpha}\biggr)+O\bigl(\veps^{2-\alpha}\bigr)
\, . \label{gman}
\end{equation}
For the given $n$ and $\alpha$, we can always choose
$\veps_{n\alpha}>0$ such that the right-hand side of (\ref{gman}) is
strictly positive for all $0<\veps<\veps_{n\alpha}\,$. In the
interval $x\in(\veps,1)$, the function
$x^{-\alpha}\varDelta_{n\alpha}(x)$ is smooth and monotone. Further,
this function is strictly positive at the left least point and zero
at the right one. By monotonicity, non-negativity takes place for
all $x\in(\veps,1)$. Making $\veps>0$ arbitrarily small completes
the proof for $\alpha\in(0,1)$. Along this proof we also conclude
that $(-1)^{n}\,\tau_{n\alpha}\geq0$ for $\alpha\in(0,1)$.

Let us proceed to the case $\alpha\in(1,2)$. It is shown in Appendix
\ref{aniq2} that $(-1)^{n}\,\tau_{n\alpha}\geq0$ for all
$\alpha\in(1,2)$. Combining this with (\ref{moden}) and
(\ref{moden1}) implies that the function
$x\mapsto{x}^{-\alpha}\varDelta_{n\alpha}(x)$ cannot increase for
all $0<x<1$. As taking zero value at $x=1$, this function is
non-negative for all points of the interval. $\blacksquare$

The derived approximation remains valid for $\alpha=2$. Taking
$\alpha=2-\epsilon$ with sufficiently small $\epsilon>0$, we have
\begin{equation}
\frac{1}{\tau_{n\alpha}}=\frac{c_{n}^{(2)}}{\epsilon}\,\bigl(1+O(\epsilon)\bigr)
\, , \label{invtau1}
\end{equation}
whence the polynomial (\ref{gnsun}) becomes
\begin{equation}
F_{n\alpha}(x)=\frac{\epsilon}{c_{n}^{(2)}}\>\frac{c_{n}^{(2)}(x^{2}-x)}{\epsilon}+O(\epsilon)
\underset{\epsilon\to0}{\longrightarrow}x^{2}-x
\, . \label{gnsun2}
\end{equation}
That is, estimation by polynomials of the form (\ref{gnsun}) is
quite exact in the case $\alpha=2$. On the other hand, we will use
estimation by polynomials (\ref{gnsun}) only for non-integer values
of $\alpha$ between $0$ and $2$.

It follows from (\ref{ineq01}) that, for
$\alpha\in(0,1)\cup(1,2)$ and $x\in[0,1]$,
\begin{equation}
\frac{x^{\alpha}-x}{1-\alpha}\geq\sum_{s=1}^{n}\wta_{n\alpha}^{(s)}\,x^{s}
\, , \label{caab1}
\end{equation}
where the coefficients
\begin{equation}
\wta_{n\alpha}^{(1)}=\tau_{n\alpha}\sum_{s=2}^{n}\frac{c_{n}^{(s)}}{s-\alpha}
\ , \qquad
\wta_{n\alpha}^{(s)}=\frac{\tau_{n\alpha}\,c_{n}^{(s)}}{\alpha-s}
\qquad
(2\leq{s}\leq{n})
\, , \label{aceff1}
\end{equation}
and $\tau_{n\alpha}$ is defined by (\ref{invtau}). Taking the limit
$\alpha\to1$ results in (\ref{yleg}). To transform (\ref{caab1})
into an estimate from above, we generalize the corresponding
derivation of \cite{rastpol}. It follows from (\ref{caab1}) that
\begin{equation}
\frac{\xdif\eta_{\alpha}(\tilde{x})}{\xdif\tilde{x}}
=\alpha\,\frac{\tilde{x}^{\alpha-1}-1}{1-\alpha}-1\geq\alpha\sum_{s=1}^{n}\wta_{n\alpha}^{(s)}\,\tilde{x}^{s-1}-1
\, . \label{difet}
\end{equation}
Integrating this from $\tilde{x}=x$ to $\tilde{x}=1$ gives
\begin{equation}
{}-\eta_{\alpha}(x)\geq\alpha\sum_{s=1}^{n}\wta_{n\alpha}^{(s)}\,\frac{1-x^{s}}{s}+x-1
\, . \label{intet}
\end{equation}
Hence, we obtain the inequality
\begin{equation}
\frac{x^{\alpha}-x}{1-\alpha}\leq\sum_{s=0}^{n}\wb_{n\alpha}^{(s)}\,x^{s}
\, , \label{baac1}
\end{equation}
where the coefficients reads as
\begin{equation}
\wb_{n\alpha}^{(0)}=1-\alpha\sum_{s=1}^{n}\frac{\wta_{n\alpha}^{(s)}}{s}
\ , \qquad
\wb_{n\alpha}^{(1)}=\alpha\,\wta_{n\alpha}^{(1)}-1
\ , \qquad
\wb_{n\alpha}^{(s)}=\frac{\alpha}{s}\>\wta_{n\alpha}^{(s)}
\qquad
(2\leq{s}\leq{n})
\, . \label{bceff1}
\end{equation}
Except for the case $s=0$, these expressions are in obvious
agreement with (\ref{bceff}). Nevertheless, the coefficient
$b_{n\alpha}^{(0)}$ can indeed be rewritten similarly to
$\wb_{n\alpha}^{(0)}$.

\section{Complementarity relations in terms of generalized entropies}\label{sec4}

In this section, we will derive entropic uncertainty and certainty
relations for POVMs assigned to a quantum design. Today, uncertainty
relations play a certain role in various fields beyond the
conceptual questions of quantum theory. The security analysis of
quantum cryptographic protocols uses information-theoretic
uncertainty relations \cite{damren,ng12}. They also give a tool to
detect non-classical correlations such as entanglement
\cite{giov04,glew04,olew1} and steerability \cite{cug18,brun18}.
Uncertainty relations were also utilized to estimate the amount of
quantum coherence \cite{xu22,rascr}. References \cite{raphyca,kong}
considered uncertainty relations in application to quantum channels.
Uncertainties in flavor and mass eigenstates of neutrinos were
addressed in \cite{shem}. Recently, entropic uncertainty relations
were used to explore the formation of quark pairs at large hadron
collider \cite{fei24}.

The entropies of Tsallis \cite{tsallis} and R\'{e}nyi \cite{renyi61} will be used to
quantify the amount of uncertainties. For the given probability
distribution $\vcp=\{p_{j}\}$ and $0<\alpha\neq1$, we write
\begin{align}
H_{\alpha}(\vcp)&:=\frac{1}{1-\alpha}\left(\sum\nolimits_{j} p_{j}^{\alpha}-1\right)
=\sum\nolimits_{j} \eta_{\alpha}(p_{j})
\ , \label{tsedf}\\
R_{\alpha}(\vcp)&:=\frac{1}{1-\alpha}\,\ln\!\left(\sum\nolimits_{j} p_{j}^{\alpha}\right)
=\frac{1}{1-\alpha}\,\ln\bigl(1+(1-\alpha)H_{\alpha}(\vcp)\bigr)
\ . \label{reedf}
\end{align}
The formulas (\ref{tsedf}) and (\ref{reedf}), respectively, define the
Tsallis and R\'{e}nyi $\alpha$-entropies. In the limit $\alpha\to1$,
each of these entropies leads to the standard Shannon entropy
\begin{equation}
H_{1}(\vcp):=-\sum\nolimits_{j} p_{j}\ln{p}_{j}
\, . \label{shdf}
\end{equation}
Basic properties of the Tsallis and R\'{e}nyi entropies are
considered in section 2.7 of \cite{bengtsson}. It will be helpful to
use the $s$-index of coincidence defined as
\begin{equation}
I^{(s)}(\vcp):=\sum\nolimits_{j} p_{j}^{s}
\, . \label{icsdef}
\end{equation}
For $s=0$, the index is equal to the number of nonzero
probabilities. Properties of such indices were briefly discussed in
\cite{harremoes}.

Suppose that we know exactly $s$-indices of coincidence for
$s=2,\ldots,n$. Then the maximal probability is estimated from above
as \cite{rastdes}
\begin{equation}
\underset{j}{\max}\,p_{j}\leq\Upsilon_{L-1}^{(n)}\bigl(I^{(n)}(\vcp)\bigr)
\, , \label{maxki}
\end{equation}
where $L$ is the number of nonzero probabilities. By
$\Upsilon_{L-1}^{(n)}\bigl(I^{(n)}(\vcp)\bigr)$, we denote the maximal
real root of the equation
\begin{equation}
(1-\Upsilon)^{n}+(L-1)^{n-1}\Upsilon^{n}=(L-1)^{n-1}I^{(n)}(\vcp)
\, . \label{curvey}
\end{equation}
This conclusion generalizes the idea originally presented in
\cite{rastmubs}. Due to (\ref{maxki}) and the results of Sect.
\ref{sec3}, the following statement takes place.

\newtheorem{tsa02}[eta02]{Proposition}
\begin{tsa02}\label{rests}
Let $I^{(s)}(\vcp)$ be given for all $s=2,\ldots,n$. For
$\alpha\in(0,2)$, the Tsallis $\alpha$-entropy satisfies
\begin{equation}
\sum_{s=1}^{n} a_{n\alpha}^{(s)}\Upsilon^{\alpha-s}I^{(s)}(\vcp)
-\Upsilon^{\alpha-1}\ln_{\alpha}(\Upsilon)
\leq{H}_{\alpha}(\vcp)\leq
\sum_{s=0}^{n} b_{n\alpha}^{(s)}\Upsilon^{\alpha-s}I^{(s)}(\vcp)
-\Upsilon^{\alpha-1}\ln_{\alpha}(\Upsilon)
\, , \label{tsotay}
\end{equation}
where $\Upsilon=\Upsilon_{L-1}^{(n)}\bigl(I^{(n)}(\vcp)\bigr)$ and
the coefficients $a_{n}^{(s)}$ and $b_{n}^{(s)}$ are defined by
(\ref{aceff}) and (\ref{bceff}), respectively. It also holds that
\begin{align}
\sum_{s=1}^{n} \wta_{n\alpha}^{(s)}\Upsilon^{\alpha-s}I^{(s)}(\vcp)
-\Upsilon^{\alpha-1}\ln_{\alpha}(\Upsilon)
\leq{H}_{\alpha}(\vcp)\leq
\sum_{s=0}^{n} \wb_{n\alpha}^{(s)}\Upsilon^{\alpha-s}I^{(s)}(\vcp)
-\Upsilon^{\alpha-1}\ln_{\alpha}(\Upsilon)
\, , \label{tsocheb}
\end{align}
where the coefficients are defined by (\ref{aceff1}) and (\ref{bceff1}).
\end{tsa02}

{\bf Proof.} The derived two-sided estimates will be used with the
points $x_{j}=p_{j}/\Upsilon$ that lie in the interval $x\in[0,1]$
due to (\ref{maxki}). Substituting these points into (\ref{tsedf})
finally leads to
\begin{equation}
H_{\alpha}(\vcp)=\Upsilon^{\alpha}\sum_{j=1}^{L}\eta_{\alpha}(x_{j})-\Upsilon^{\alpha-1}\ln_{\alpha}(\Upsilon)
\, . \label{haux}
\end{equation}
Applying (\ref{yaab1}) and (\ref{ybab1}) to each $x_{j}$, the
summation gives
\begin{equation}
\sum_{s=1}^{n} a_{n\alpha}^{(s)}\Upsilon^{-s}I^{(s)}(\vcp)
\leq\sum_{j=1}^{L}\eta_{\alpha}(x_{j})\leq
\sum_{s=0}^{n} b_{n\alpha}^{(s)}\Upsilon^{-s}I^{(s)}(\vcp)
\ . \label{tayty}
\end{equation}
Combining (\ref{haux}) with (\ref{tayty}) completes the proof of
(\ref{tsotay}). Similarly, the two-sided estimate (\ref{tsocheb})
follows from (\ref{caab1}) and (\ref{baac1}). $\blacksquare$

The statement of Proposition \ref{rests} provides two-sided
estimates on the Tsallis $\alpha$-entropy for $\alpha\in(0,2)$ in
terms of several indices of integer degree. It generalizes two-sided
estimates on the Shannon entropy proposed in \cite{rastpol}. The
two-sided estimates (\ref{tsotay}) and (\ref{tsocheb}) are easily
adopted to the case of R\'{e}nyi entropies. For $0<\alpha\neq1$, the
function $z\mapsto(1-\alpha)^{-1}\ln\bigl(1+(1-\alpha)z\bigr)$
increases. Combining this with (\ref{reedf}) gives
\begin{equation}
\frac{1}{1-\alpha}\,\ln\bigl(1+(1-\alpha)A_{n\alpha}(\vcp)\bigr)
\leq{R}_{\alpha}(\vcp)
\leq\frac{1}{1-\alpha}\,\ln\bigl(1+(1-\alpha)B_{n\alpha}(\vcp)\bigr)
\, , \label{retych}
\end{equation}
whenever
$A_{n\alpha}(\vcp)\leq{H}_{\alpha}(\vcp)\leq{B}_{n\alpha}(\vcp)$.
Here, the term $A_{n\alpha}(\vcp)$ stands for the left-hand side of
(\ref{tsotay}) or (\ref{tsocheb}), and the term $B_{n\alpha}(\vcp)$
stands for the right-hand side of (\ref{tsotay}) or (\ref{tsocheb}).
For the sake of shortness, we refrain from presenting the explicit
formulas.

By $H_{\alpha}(\cle^{(m)};\bro)$ and $R_{\alpha}(\cle^{(m)};\bro)$,
we, respectively, denote the $\alpha$-entropies (\ref{tsedf}) and
(\ref{reedf}) calculated with probabilities (\ref{prbk}). The
results (\ref{tsotay}) or (\ref{tsocheb}) immediately lead to
complementarity relations for POVMs assigned to a quantum design.
The formulation in terms of Tsallis entropies is posed as follows.

\newtheorem{tsads}[eta02]{Proposition}
\begin{tsads}\label{res2}
Let $M$ rank-one POVMs $\cle^{(m)}$, each with $\ell$ elements of
the form (\ref{mejdf}), be assigned to a quantum $t$-design
$\dn=\bigl\{|\phi_{k}\rangle\bigr\}_{k=1}^{K}$ in $d$ dimensions.
For $\alpha\in(0,2)$, it then holds that
\begin{align}
\sum_{s=1}^{t} a_{t\alpha}^{(s)}\Upsilon^{\alpha-s}\bar{\beta}_{\ell}^{(s)}(\bro)
-\Upsilon^{\alpha-1}\ln_{\alpha}(\Upsilon)
&\leq\frac{1}{M}\sum_{m=1}^{M}H_{\alpha}(\cle^{(m)};\bro)
\leq\sum_{s=0}^{t} b_{t\alpha}^{(s)}\Upsilon^{\alpha-s}\bar{\beta}_{\ell}^{(s)}(\bro)
-\Upsilon^{\alpha-1}\ln_{\alpha}(\Upsilon)
\, , \label{apotay}\\
\sum_{s=1}^{t} \wta_{t\alpha}^{(s)}\Upsilon^{\alpha-s}\bar{\beta}_{\ell}^{(s)}(\bro)
-\Upsilon^{\alpha-1}\ln_{\alpha}(\Upsilon)
&\leq\frac{1}{M}\sum_{m=1}^{M}H_{\alpha}(\cle^{(m)};\bro)
\leq\sum_{s=0}^{t} \wb_{t\alpha}^{(s)}\Upsilon^{\alpha-s}\bar{\beta}_{\ell}^{(s)}(\bro)
-\Upsilon^{\alpha-1}\ln_{\alpha}(\Upsilon)
\, , \label{apoche}
\end{align}
where $\Upsilon=\min\bigl\{M\,\Upsilon_{K-1}^{(t)}\bigl(\bar{\beta}^{(t)}(\bro)\bigr),1\bigr\}$.
\end{tsads}

{\bf Proof.} The left-hand side of (\ref{mindt}) is equal to the sum
of $s$-indices over all POVMs. It follows from (\ref{prbk}) and
(\ref{maxki}) that \cite{rastpol}
\begin{equation}
\frac{d}{\ell}\,\langle\phi_{j}^{(m)}|\bro|\phi_{j}^{(m)}\rangle
=p_{j}(\cle^{(m)};\bro)\leq{M}\,\Upsilon_{K-1}^{(t)}\bigl(\bar{\beta}^{(t)}(\bro)\bigr)
\, , \label{depm}
\end{equation}
where $K=\ell{M}$. For each of $M$ entropies
$H_{\alpha}(\cle^{(m)};\bro)$, we use (\ref{tsotay}) with maximal
power $t$ and the taken $\Upsilon$, whence
\begin{equation}
\sum_{s=1}^{t} a_{t\alpha}^{(s)}\Upsilon^{\alpha-s}I^{(s)}(\vcp^{(m)})
-\Upsilon^{\alpha-1}\ln_{\alpha}(\Upsilon)
\leq{H}_{\alpha}(\cle^{(m)};\bro)
\leq\sum_{s=0}^{t} b_{t\alpha}^{(s)}\Upsilon^{\alpha-s}I^{(s)}(\vcp^{(m)})
-\Upsilon^{\alpha-1}\ln_{\alpha}(\Upsilon)
\, . \label{potays}
\end{equation}
It is seen from (\ref{mindt}) that, for all $s=2,\ldots,t$,
\begin{equation}
\sum_{m=1}^{M}I^{(s)}(\vcp^{(m)})=
K\ell^{-s}d^{\,s}\,\cald_{d}^{(s)}\,\tr\bigl(\bro^{\otimes{s}}\psyms\bigr)=
M\bar{\beta}_{\ell}^{(s)}(\bro)
\, . \label{sindet}
\end{equation}
Summing (\ref{potays}) with respect to $m$ and substituting
(\ref{sindet}), we obtain (\ref{apotay}) multiplied by factor $M$.
In a similar manner, we derive (\ref{apoche}) from (\ref{tsocheb}).
$\blacksquare$

For $\alpha\in(0,2)$, we have two families of complementarity
relations for the Tsallis $\alpha$-entropy averaged over all $M$
POVMs $\cle^{(m)}$. When single POVM $\cle$ with $K$ elements
(\ref{mkdef}) is assigned to the given $t$-design, the relations
(\ref{apotay}) and (\ref{apoche}), respectively, reduce to
\begin{align}
\sum_{s=1}^{t} a_{t\alpha}^{(s)}\Upsilon^{\alpha-s}\bar{\beta}^{(s)}(\bro)
-\Upsilon^{\alpha-1}\ln_{\alpha}(\Upsilon)
&\leq{H}_{\alpha}(\cle;\bro)\leq
\sum_{s=0}^{t} b_{t\alpha}^{(s)}\Upsilon^{\alpha-s}\bar{\beta}^{(s)}(\bro)
-\Upsilon^{\alpha-1}\ln_{\alpha}(\Upsilon)
\, , \label{kapotay}\\
\sum_{s=1}^{t}\wta_{t\alpha}^{(s)}\Upsilon^{\alpha-s}\bar{\beta}^{(s)}(\bro)
-\Upsilon^{\alpha-1}\ln_{\alpha}(\Upsilon)
&\leq{H}_{\alpha}(\cle;\bro)
\leq\sum_{s=0}^{t}\wb_{t\alpha}^{(s)}\Upsilon^{\alpha-s}\bar{\beta}^{(s)}(\bro)
-\Upsilon^{\alpha-1}\ln_{\alpha}(\Upsilon)
\, . \label{kapoche}
\end{align}
The complementarity relations (\ref{apotay}), (\ref{apoche}),
(\ref{kapotay}), (\ref{kapoche}) are further combined with
(\ref{retych}). In this way, one can write R\'{e}nyi formulation of
complementarity relations for $\alpha\in(0,2)$. We refrain from
presenting the details here. For a pure state, the formulas
(\ref{kapotay}) and (\ref{kapoche}) read as
\begin{align}
H_{\alpha}\bigl(\cle;|\psi\rangle\langle\psi|\bigr)
&\geq\sum_{s=1}^{t} a_{t\alpha}^{(s)}\Upsilon^{\alpha-s}K^{1-s}d^{\,s}\,\cald_{d}^{(s)}
-\Upsilon^{\alpha-1}\ln_{\alpha}(\Upsilon)
\, , \label{ppit}\\
H_{\alpha}\bigl(\cle;|\psi\rangle\langle\psi|\bigr)
&\geq\sum_{s=1}^{t}\wta_{t\alpha}^{(s)}\Upsilon^{\alpha-s}K^{1-s}d^{\,s}\,\cald_{d}^{(s)}
-\Upsilon^{\alpha-1}\ln_{\alpha}(\Upsilon)
\, , \label{ppich}
\end{align}
where
$\Upsilon=\Upsilon_{K-1}^{(t)}\bigl(K^{1-t}d^{\,t}\,\cald_{d}^{(t)}\bigr)$.
The following fact will be seen with concrete examples of quantum
designs. The lower entropic bounds obtained for pure states remain
valid for all states. This conclusion seems to be physically
natural, but its validity for (\ref{ppit}) and (\ref{ppich}) is not
easy to prove analytically. At the same time, the claim can always
be checked by numerical inspection in each concrete example of
design-structured POVMs \cite{rastpol}. If is true, the right-hand
sides of (\ref{ppit}) and (\ref{ppich}) give a state-independent
formulation of uncertainty relations.

In the following, we consider several examples to inspect our
findings. For $\alpha=1$, they reduce to complementarity relations
in terms of the Shannon entropy. Such relations were proposed and
exemplified in \cite{rastpol}. It was found therein that the
Shannon-entropy relations provide better results for single POVM
assigned to a quantum design. In general, the same conclusion holds
for complementarity relations in terms of $\alpha$-entropies of
order $\alpha\in(0,2)$. It origins in the fact that, for $M>1$, we
sometimes deal with the trivial substitution $\Upsilon=1$. For this
reason, we will mainly focus on (\ref{kapotay}) and (\ref{kapoche})
concerning the case of single assigned POVM. There are interesting
examples of quantum designs in two dimensions, including cases with
sufficiently large $K$. Such examples are more sensible to
illustrate features of the proposed complementarity relations.

For assessment of the new results, we also use uncertainty relations
in terms of Tsallis entropies derived in \cite{rastrid}. At the
given index of coincidence, these inequalities estimate entropies
from below for $\alpha\in(0,2]$. Let the piecewise smooth function
$x\mapsto{L}_{\alpha}(x)$ be defined as
\begin{equation}
L_{\alpha}(x)=(q+1)\ln_{\alpha}(q+1)-q\ln_{\alpha}(q)
-q(q+1)\bigl[\,\ln_{\alpha}(q+1)-\ln_{\alpha}(q)\bigr]x
\, , \qquad
x\in\biggl[\frac{1}{q+1}\, ,\frac{1}{q}\biggr]
\, , \label{elaq}
\end{equation}
where integer $q\geq1$. For $\alpha\in(0,2]$, it holds that \cite{rastrid}
\begin{equation}
H_{\alpha}(\vcp)\geq{L}_{\alpha}\bigl(I^{(2)}(\vcp)\bigr)
\, . \label{rideq}
\end{equation}
For $\alpha=1$, this inequality reduces to one of the main results
proved in \cite{harremoes}. For a single POVM assigned to a quantum
design, we then have
\begin{equation}
H_{\alpha}(\cle;\bro)\geq
L_{\alpha}\bigl(\bar{\beta}^{(2)}(\bro)\bigr)
\, . \label{ridde}
\end{equation}
Inequalities of the form (\ref{ridde}) will be used for comparison
and assessment of new uncertainty relations formulated in
Proposition \ref{res2}.

The authors of \cite{guhne20} gave a Bloch-sphere description of
several quantum designs in two dimensions. These examples are
immediately inspired by spherical designs in three-dimensional real
space examined in \cite{hardin96}. Each of used Bloch vectors comes
to one of vertices of certain polyhedron. The qubit density matrix
will be characterized by its minimal eigenvalue $\lambda$. It is
enlightening to visualize the complementarity relations
(\ref{kapotay}) and (\ref{kapoche}). To avoid bulky legends on
pictures, the following notation will be used. By ``LT-estimate''
and ``UT-estimate,'' we mean the left- and right-hand sides of
(\ref{kapotay}), respectively. They approximate the function of
interest by polynomials with coefficients according to the Taylor
scheme. The terms ``LCh-estimate'' and ``UCh-estimate,'' respectively,
refer to the left- and right-hand sides of (\ref{kapoche}). These
estimates use polynomials with flexible coefficients. The right-hand
side of (\ref{ridde}) will be referred to as
``$L_{\alpha}$-estimate.''

\begin{figure*}
\centering \includegraphics[height=7.6cm]{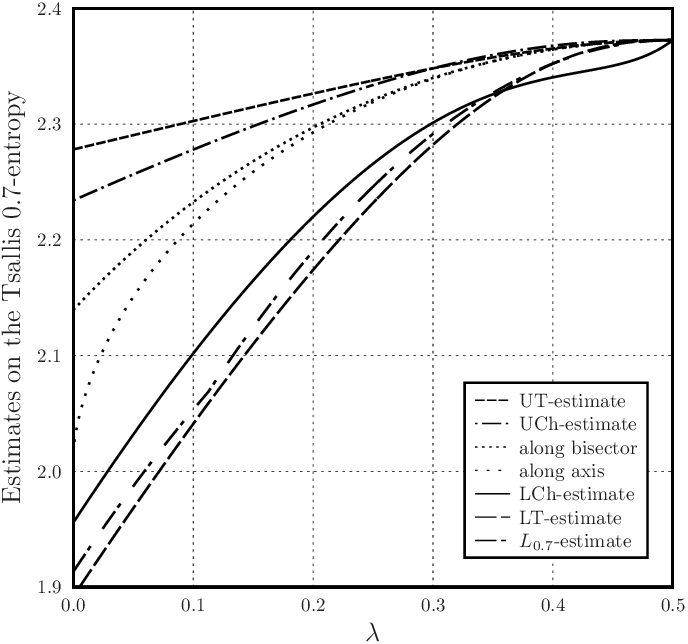}
\caption{\label{fig0} Estimates on the Tsallis $0.7$-entropy versus $\lambda$ for the $3$-design with $6$ vertices.}
\end{figure*}

For reasons that will become clear a little later, we shall start
with a value of the range $\alpha\in(0,1)$, say, $\alpha=0.7$. Let
us take the $3$-design with $K=6$ vertices of octahedron. One of
reasons to take this design is that it is formed by eigenstates of
the Pauli matrices. These vectors constitute three mutually unbiased
bases which are often used to exemplify general relations. The five
estimates as functions of $\lambda$ are shown in Fig. \ref{fig0}.
The two-sided estimate (\ref{kapotay}) is better only for states
sufficiently close to the maximally mixed one. Overall, relative
differences of two-sided estimating are no more than $12.5$ \%. It
is interesting that there is a little range of $\lambda$ in which
$L_{\alpha}$-estimate is better that both ``LCh-estimate'' and
``LT-estimate.'' In relative scale, however, this advantage is
around few thousandths. By two dotted lines, we also show values of
the $0.7$-entropy for two orientations of the Bloch vector. Let the
three coordinate axes pass through vertices of octahedron. One of
the used orientations is along a coordinate axis, whereas the other
lies in a coordinate plane along quadrant bisector. Near the right
least point $\lambda=1/2$, all the curves converge at one point.

\begin{figure*}
\centering \includegraphics[height=7.6cm]{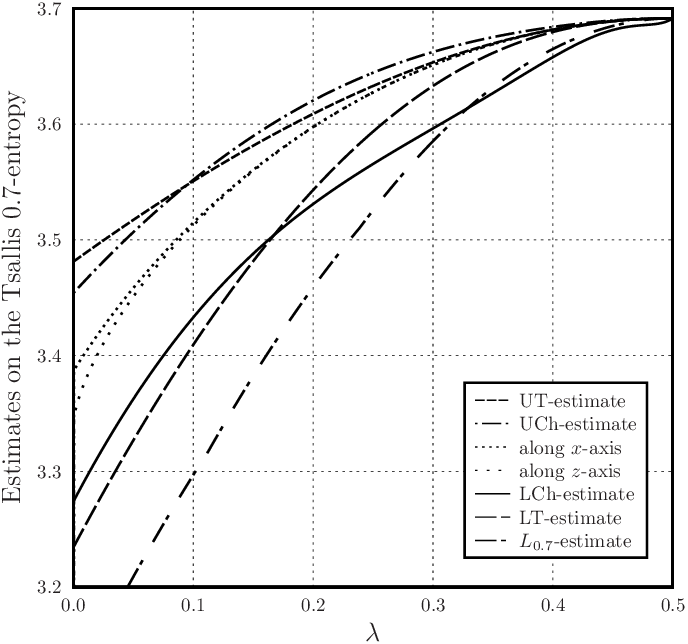}
\caption{\label{fig1} Estimates on the Tsallis $0.7$-entropy versus $\lambda$ for the $5$-design with $12$ vertices.}
\end{figure*}

Further, we consider the $5$-design with $K=12$ vertices forming an
icosahedron. In Fig. \ref{fig1}, we plot the three lower estimates
and the two upper ones as functions of $\lambda$. The two-sided
estimate (\ref{kapoche}) is better for states of moderate mixedness.
Moreover, we see a range, where ``LCh-estimate'' is sufficiently poor in
comparison with ``LT-estimate. For the Shannon entropies, ``LCh-estimate''
and ``LT-estimate'' do not differ so essentially \cite{rastpol}.
Nevertheless, one still has a wide domain, in which ``LCh-estimate'' is
better. Relative differences of two-sided estimating do not exceed
$5.2$ \%. Except for the right least point $\lambda=1/2$,
``$L_{\alpha}$-estimate'' is weaker than the maximum of
``LCh-estimate'' and ``LT-estimate.'' For the $5$-design, more
restrictions imposed than for the $3$-design. By two dotted lines,
we also show values of the $0.7$-entropy for two orientations of the
Bloch vector. Let the $z$-axis pass through two opposite vertices
and form symmetry axis of icosahedron. Let the $x$-axis pass so that
one of inclined edges lies in the $zx$-plane. One of the used
orientations is along the $z$-axis, whereas the other is along
positive part of the $x$-axis. For the maximally mixed state, all
the curves converge at one point.

\begin{figure*}
\centering \includegraphics[height=7.6cm]{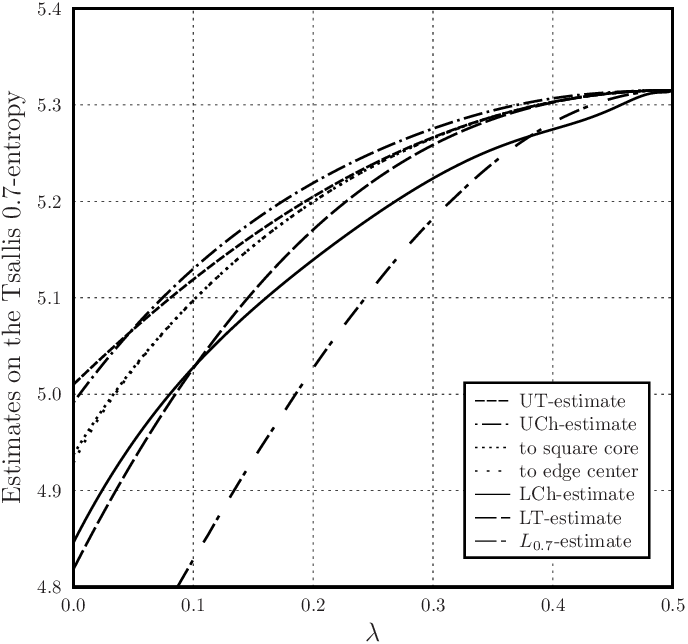}
\caption{\label{fig2} Estimates on the Tsallis $0.7$-entropy versus $\lambda$ for the $7$-design with $24$ vertices.}
\end{figure*}

Further, we recall the $7$-design with $K=24$ vertices of deformed
snub cube as described in section 3 of \cite{hardin96}. Its faces
contains $6$ squares and $32$ triangles. In Fig. \ref{fig2}, we plot
the estimates of interest versus $\lambda$. The two-sided estimate
(\ref{kapoche}) is better only for states of low mixedness. There is
again a range, where ``LCh-estimate'' is relatively poor. On the other
hand, both the ``LCh- and UCh-estimates'' are still better for pure
states. It is well known that bounds for pure states play an
important role in deriving separability criteria and steering
inequalities. In details, we will address this question later.
Relative differences of two-sided estimating are no greater than
$2.9$ \%. Similarly to Fig. \ref{fig1}, ``$L_{\alpha}$-estimate'' is
far from advantage. By two dotted lines, we also show values of the
$0.7$-entropy for two orientations of the Bloch vector. The first
direction goes to the square core, whereas the other goes to center
of of one of square edges. In this example, the two dotted lines are
closer than in Fig. \ref{fig1}. All six curves tend to coincide on
the right. Distinctions between them are less than in other
examples.

The above examples allow us to make several conclusions. We see that
the ``LCh- and UCh-estimates'' provide better results for pure states
and states close to them. For states of sufficient mixedness, the
``LT- and UT-estimates'' should be preferred. With growth of degree $t$,
more restrictions are actually imposed on the probabilities. Their
distribution becomes more uniform, though number of outcomes also
grows. This fact leads to shifting points of approximation closer to
the right point $x=1$. So, we observe increasing role of the
estimates with coefficients according to the Taylor scheme.
Nevertheless, the estimates with flexible coefficients still remain
better for pure states. It is well known that pure-state entropic
bounds are significant to derive separability and steering criteria.
Hence, the ``LCh- and UCh-estimates'' are of interest, even if domains
of their dominance shorten. Except for the $3$-design,
``$L_{\alpha}$-estimate'' cannot provide an advantage for this value
of $\alpha$. In the next section, we will consider separability
criteria and steering inequalities in more detail.

\begin{figure*}
\centering \includegraphics[height=7.6cm]{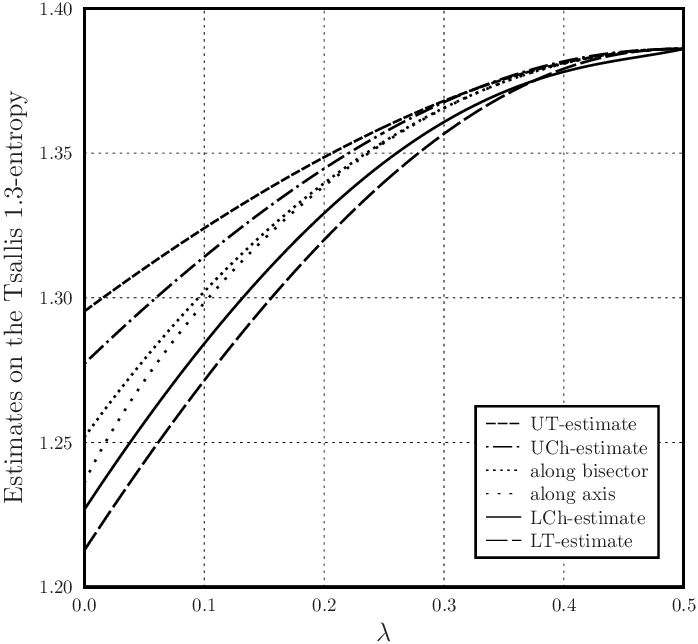}
\caption{\label{fig3} Estimates on the Tsallis $1.3$-entropy versus $\lambda$ for the $3$-design with $6$ vertices.}
\end{figure*}

Finally, we briefly exemplify the derived estimates for a value of
the range $\alpha\in(1,2)$. In Fig. \ref{fig3}, the four estimates
from (\ref{kapotay}) and (\ref{kapoche}) are plotted for
$\alpha=1.3$ together with the actual entropies for two orientations
of the Bloch vector. ``$L_{\alpha}$-estimate'' is not shown since it
goes very close to ``LT-estimate''. In contrast to Fig. \ref{fig0},
there is no advantage of ``$L_{\alpha}$-estimate'' over the maximum
of ``LCh-estimate'' and ``LT-estimate''. Here, relative differences
of two-sided estimating are no more than $3.9$ \%. Overall, both the
approaches to estimate $\alpha$-entropies demonstrate domains of
utility, though the estimate (\ref{kapoche}) is better only for
states very close to the maximally mixed one. Entropies themselves
and estimates on them do not differ so essentially as in Fig.
\ref{fig0}. In that figure, the graph of ``LCh-estimate'' shows a fairly
convex part at the right end of the interval. In Fig. \ref{fig3},
this behavior is not expressed so brightly. By some inspection,
similar observations can be made for other examples of quantum
designs in two dimensions. We refrain from presenting the details
here.

By comparison, the two approaches to estimate $\alpha$-entropies are
more different for $\alpha\in(0,1)$. Indeed, the first derivative of
$\eta_{\alpha}(x)$ is not finite for $x\to+0$ and $\alpha\in(0,1)$.
On the other hand, derivatives of any polynomial are certainly
finite, whence the slope of tangential line is enough far from the
vertical. For $\alpha\in(0,1)$, therefore, our approximation have to
be relatively poor in some right neighborhood of the left point
$x=0$. By its origin, the Taylor scheme provides the best way around
the right point $x=1$. The polynomials with flexible coefficients
give better results closely to the left point of the interval.
Similar observations were reported for the Shannon entropy
\cite{rastpol}. For $\alpha\in(1,2)$, the first derivative of
$\eta_{\alpha}(x)$ remains finite for $x\to+0$, so that both the
schemes allow one to estimate $\alpha$-entropy in a good way. As some
inspection shows, we should prefer (\ref{kapotay}) for sufficiently
small values of $\alpha$. For non-integer values $\alpha>2$, one can
use the methods already presented in the literature. We refrain from
presenting the details here.

\section{Some applications of the main results}\label{sec5}

In this section, we consider applications of the derived relations
for tasks of quantum information. It was discussed in \cite{rastpol}
how to use design-structured POVMs for estimating the von Neumann
entropy of a quantum state. Due to (\ref{kapotay}) and
(\ref{kapoche}), the same measurements can be used for estimating
generalized quantum entropies. To the given density matrix $\bro$
and $0<\alpha\neq1$, we assign the Tsallis $\alpha$-entropy
\begin{equation}
\rmH_{\alpha}(\bro):=\frac{\tr(\bro^{\alpha})-1}{1-\alpha}
\ . \label{atsab}
\end{equation}
The quantum R\'{e}nyi $\alpha$-entropy reads as
\begin{equation}
\rmR_{\alpha}(\bro):=\frac{\ln\bigl(\tr(\bro^{\alpha})\bigr)}{1-\alpha}=
\frac{1}{1-\alpha}\,\ln\bigl(1+(1-\alpha)\rmH_{\alpha}(\bro)\bigr)
\, . \label{areab}
\end{equation}
In the limit $\alpha\to1$, both the above formulas gives the von
Neumann entropy $\rmH_{1}(\bro)=\!{}-\tr(\bro\ln\bro)$. Using
quantities of the form (\ref{betn}) or (\ref{betk}), one can
recalculate the traces $\tr(\bro^{s})$ for $s=2,\ldots,t$. Applying
(\ref{kapotay}) and (\ref{kapoche}) to this case results in
\begin{align}
\sum_{s=1}^{t} a_{t\alpha}^{(s)}\Lambda^{\alpha-s}\,\tr(\bro^{s})
-\Lambda^{\alpha-1}\ln_{\alpha}(\Lambda)
&\leq\rmH_{\alpha}(\bro)\leq{b}_{t\alpha}^{(0)}\Lambda^{\alpha}d+
\sum_{s=1}^{t} b_{t\alpha}^{(s)}\Lambda^{\alpha-s}\,\tr(\bro^{s})
-\Lambda^{\alpha-1}\ln_{\alpha}(\Lambda)
\, , \label{potayka}\\
\sum_{s=1}^{t}\wta_{t\alpha}^{(s)}\Lambda^{\alpha-s}\,\tr(\bro^{s})
-\Lambda^{\alpha-1}\ln_{\alpha}(\Lambda)
&\leq\rmH_{\alpha}(\bro)
\leq\wb_{t\alpha}^{(0)}\Lambda^{\alpha}d+
\sum_{s=1}^{t}\wb_{t\alpha}^{(s)}\Lambda^{\alpha-s}\,\tr(\bro^{s})
-\Lambda^{\alpha-1}\ln_{\alpha}(\Lambda)
\, , \label{pocheka}
\end{align}
where $\alpha\in(0,2)$ and
$\Lambda=\Upsilon_{d-1}^{(t)}\bigl(\tr(\bro^{\!\;t})\bigr)$. The latter
estimates the eigenvalues of $\bro$ from above \cite{rastpol}.
Similarly to (\ref{retych}), we have
\begin{equation}
\frac{1}{1-\alpha}\,\ln\bigl(1+(1-\alpha)\rmA_{n\alpha}(\bro)\bigr)
\leq\rmR_{\alpha}(\bro)
\leq\frac{1}{1-\alpha}\,\ln\bigl(1+(1-\alpha)\rmB_{n\alpha}(\bro)\bigr)
\, , \label{ertych}
\end{equation}
whenever
$\rmA_{n\alpha}(\bro)\leq\rmH_{\alpha}(\bro)\leq\rmB_{n\alpha}(\bro)$.
Here, the term $\rmA_{n\alpha}(\bro)$ stands for the left-hand side
of (\ref{potayka}) or (\ref{pocheka}), and the term
$\rmB_{n\alpha}(\bro)$ stands for the right-hand side of
(\ref{potayka}) or (\ref{pocheka}). In this way, the quantum
R\'{e}nyi $\alpha$-entropy of order $\alpha\in(0,2)$ can be
estimated.

The phenomenon of entanglement attracts an attention since the
Schr\"{o}dinger ``cat paradox'' paper \cite{erwin1935} and in the
Einstein--Podolsky--Rosen paper \cite{epr1935} were published. Also,
it is an indispensable tool for quantum information processing
\cite{hhhh09}. Many separability conditions follow from uncertainty
relations \cite{glew04,giov04,devs05,gmta06}. A bipartite mixed
state is called separable, when its density matrix can be
represented as a convex combination of product states
\cite{werner89,zhsl98}. To derive separability conditions from
(\ref{kapotay}) and (\ref{kapoche}), we recall the scheme proposed
in \cite{rastsep}. Let $\cln_{A}=\{\nm_{Aj}\}$ and
$\cln_{B}=\{\nm_{Bk}\}$ be two $K$-outcome POVMs in $\hh_{A}$ and
$\hh_{B}$, respectively. It is helpful to numerate POVM elements by
numbers from $0$ up to $K-1$. To the given POVMs, we assign the
measurement $\clm(\cln_{A},\cln_{B})$ with $K$ elements
\begin{equation}
\mn_{k}:=\sum_{j=0}^{K-1}   \nm_{Aj}\otimes\nm_{Bk\ominus{j}}
\, , \label{mnk}
\end{equation}
where the symbol $\ominus$ means the modular subtraction. It follows
from (\ref{mnk}) that \cite{rastsep}
\begin{equation}
\vcp\bigl(\clm;\bro_{A}\otimes\bro_{B}\bigr)=\vcp(\cln_{A};\bro_{A})*\vcp(\cln_{B};\bro_{B})
\, . \label{mnk1}
\end{equation}
That is, for product states one generates the convolution of two
distributions assigned to local measurements. Hence, the
distribution (\ref{mnk1}) is majorized by any of two local
probability distributions \cite{rastsep}. Let one of two POVMs
$\cln_{A}$ and $\cln_{B}$, say $\cln_{A}$, be assigned to a quantum
design in line with (\ref{mkdef}). To get separability conditions,
we need state-independent uncertainty bounds. As was already noted,
validity of the lower entropic bounds (\ref{ppit}) and (\ref{ppich})
for all states should be checked in each concrete example. If they
really give a state-independent formulation, then we have the
following result. For each separable state $\bro_{AB}$ and
$\alpha\in(0,2)$, it holds that
\begin{align}
H_{\alpha}(\clm;\bro_{AB})
&\geq\sum_{s=1}^{t} a_{t\alpha}^{(s)}\Upsilon^{\alpha-s}K^{1-s}d^{\,s}\,\cald_{d}^{(s)}
-\Upsilon^{\alpha-1}\ln_{\alpha}(\Upsilon)
\, , \label{psit}\\
H_{\alpha}(\clm;\bro_{AB})
&\geq\sum_{s=1}^{t} \wta_{t\alpha}^{(s)}\Upsilon^{\alpha-s}K^{1-s}d^{\,s}\,\cald_{d}^{(s)}
-\Upsilon^{\alpha-1}\ln_{\alpha}(\Upsilon)
\, , \label{psich}
\end{align}
where $d=\dim(\hh_{A})$ and
$\Upsilon=\Upsilon_{K-1}^{(t)}\bigl(K^{1-t}d^{\,t}\,\cald_{d}^{(t)}\bigr)$.
The separability criteria (\ref{psit}) and (\ref{psich}),
respectively, follow from (\ref{ppit}) and (\ref{ppich}) by
Schur-concavity and concavity of the Tsallis $\alpha$-entropy. The
proof is carried out quite similarly to proposition 8 of the paper
\cite{rastsep}. Another  way is based on uncertainty relations of
the Landau--Pollak type \cite{devs05}. It follows from the results
of \cite{rastdes} that
\begin{equation}
\underset{k}{\max}\,p_{k}(\cln_{A};\bro_{A})
\leq\Upsilon_{K-1}^{(t)}\bigl(\bar{\beta}^{(t)}(\bro_{A})\bigr)\leq\Upsilon_{K-1}^{(t)}\bigl(K^{1-t}d^{\,t}\,\cald_{d}^{(t)}\bigr)
\, . \label{pmak}
\end{equation}
That is, we {\it a priori} know estimating from above that holds for
all states. For any separable state $\bro_{AB}$, therefore, we have
\begin{equation}
\underset{k}{\max}\,p_{k}(\clm;\bro_{AB})\leq\Upsilon_{K-1}^{(t)}\bigl(K^{1-t}d^{\,t}\,\cald_{d}^{(t)}\bigr)
\, . \label{pmak1}
\end{equation}
The above separability criteria mirror conditions imposed on
measurement statistics in the case of a single POVM. Quantum designs
also give a tool to characterize multipartite entanglement
\cite{kwg20}. Applications of the derived complementarity relations
to multipartite systems deserve further investigations.

The question of using quantum designs to derive quantum steering
inequalities was addressed in \cite{guhne20,rastdes,rastpol}.
Quantum steering inequalities are currently the subject of active
research. For more details, see the reviews \cite{cs2017,ucno20} and
references therein. It is natural that steering criteria are closely
linked to uncertainty relations. Steering inequalities in terms of
Shannon entropies were considered in \cite{sbwch13,kdr15,rmm18}.
Steering criteria also follow from uncertainty relations using
generalized entropies, including both the Tsallis \cite{cug18} and
R\'{e}nyi entropies \cite{brun18}. The novel uncertainty relations
presented in this paper can immediately be incorporated into the
frameworks developed in \cite{cug18,brun18}. Here, we refrain from
delving into details of this multifaceted issue.

Finally, we consider the case of detection inefficiencies, when the
``no-click'' event sometimes happens \cite{rastmubs}. To the given
detector efficiency $\vark\in[0.5,1]$ and distribution
$\vcp=\{p_{j}\}$, we assign a ``distorted'' distribution
$\vcp^{(\vark)}$ such that
\begin{equation}
p_{j}^{(\vark)}=\vark\, p_{j}
\, , \qquad
p_{\varnothing}^{(\vark)}=1-\vark
\, . \label{petad}
\end{equation}
The probability $p_{\varnothing}^{(\vark)}$ characterizes the
no-click event. Indices of the form (\ref{icsdef}) are then altered
as
\begin{equation}
I^{(s)}\bigl(\vcp^{(\vark)}\bigr)=\vark^{s}I^{(s)}(\vcp)+(1-\vark)^{s}
\, . \label{alsdef}
\end{equation}
Altered indices are really dealt with in applications of the
presented complementarity relations. So, a special attention should
be paid to determining actual $\vark$ as accurately as possible.
Substituting this value, the original indices $I^{(s)}(\vcp)$ are
extracted due to (\ref{alsdef}). After that, we are ready to apply
the complementarity relations in practical questions of quantum
information processing. Of course, used detectors must have
sufficiently high efficiency.

\section{Conclusions}\label{sec6}

We have obtained complementarity relations for POVMs assigned to a
quantum design. To express the amount of uncertainties
quantitatively, the R\'{e}nyi and Tsallis entropies of order
$\alpha\in(0,2)$ were used. The two methods to estimate functions of
interest were utilized. The first method is based on truncated
expansions of the Taylor type. The second way uses polynomials with
flexible coefficients. Within applied analysis, such expansions were
motivated and illustrated by Lanczos \cite{lanczos}. Both the methods
lead to two-sided estimates on the Tsallis and R\'{e}nyi entropies,
whenever several indices of subsequent integer degrees are
prescribed. Design-structured POVMs are an obvious example in which
this situation occurs. Hence, the complementarity relations for
related measurements in terms of generalized entropies of order
$\alpha\in(0,2)$ are naturally. They give a natural extension of the
main results of \cite{rastpol}. The derived two-sided estimates are
useful in any context, where the sums of certain powers of
probabilities are easy to obtain.

The presented complementarity relations were exemplified with
several quantum designs in two dimensions. It was observed that the
two methods of estimating are more different for $\alpha\in(0,1)$.
With growth of degree $t$, we see a reduction of the domain, where
the second way is better. On the other hand, the scheme of
polynomials with flexible coefficients remains important to estimate
$\alpha$-entropies for a pure state. As is well known, pure-state
entropic bounds are used to derive separability criteria and
steering inequalities. Hence, the second scheme to formulate
two-sided estimates will be interesting, even if a domain of its
role shortens. Possible applications of new relations in quantum
tomography and entanglement detection were illustrated. As was also
mentioned, novel entropic steering inequalities follow from the
derived relations. All these results newly show a utility of
polynomials with flexible coefficients initially proposed by Lanczos
\cite{lanczos}.

\appendix

\section{An inequality}\label{aniq2}

Let us consider the polynomial function of $x\in[0,1]$ defined as
\begin{equation}
\phi_{n}(x)=\sum_{s=2}^{n}
\frac{c_{n}^{(s)}x^{s}}{s-1}
\ , \label{pdef}
\end{equation}
where $n\geq2$. It is immediate to obtain
\begin{equation}
\frac{\xdif^{2}\phi_{n}(x)}{\xdif{x}^{2}}=\sum_{s=2}^{n}
s\,c_{n}^{(s)}x^{s-2}=\frac{1}{x}\left(
\frac{\xdif{T}_{n}^{*}(x)}{\xdif{x}}
-c_{n}^{(1)}\right)
 . \label{sdp}
\end{equation}
In terms of $\xi=2x-1=\cos\theta$ with $\theta\in[0,\pi]$, we have
$T_{n}^{*}(x)=T_{n}(\cos\theta)=\cos{n}\theta$ and
\begin{equation}
\frac{\xdif{T}_{n}^{*}(x)}{\xdif{x}}=   \frac{2\,\xdif{T}_{n}(\xi)}{\xdif\xi}=
\frac{2\,\xdif\theta}{\xdif\cos\theta}\,\frac{\xdif\cos{n}\theta}{\xdif\theta}=
\frac{2n\sin{n}\theta}{\sin\theta}
\ . \label{rsin}
\end{equation}
Combining this with $c_{n}^{(1)}=(-1)^{n+1}2n^{2}$ and (\ref{sdp})
leads to
\begin{equation}
\frac{x\sin\theta}{2n}\>\frac{\xdif^{2}\phi_{n}(x)}{\xdif{x}^{2}}=(-1)^{n}
\Bigl[n\sin\theta+(-1)^{n}\sin{n}\theta\Bigr]
\, . \label{xsd2p}
\end{equation}
It can be shown by induction that, for $\theta\in[0,\pi]$ and integer $n\geq1$,
\begin{equation}
\left|\sin{n}\theta\right|\leq{n}\sin\theta
\, . \label{sinn0}
\end{equation}
In view of (\ref{sinn0}), the sign of (\ref{xsd2p}) is determined by
$(-1)^{n}$. For $x\in[0,1]$, therefore, the function $\phi_{n}(x)$
is convex for even $n$ and concave for odd $n$. Hence, we also see
convexity of the right-hand side of (\ref{yleg}). Due to
$\phi_{n}(0)=0$ and $\left.\xdif\phi_{n}/\xdif{x}\,\right|_{x=0}=0$,
the tangent line is horizontal at the point $x=0$ and coincides with
the abscissa. Combining this with convexity and concavity,
respectively, one gets $\phi_{n}(x)\geq0$ for even $n$ and
$\phi_{n}(x)\leq0$ for odd $n$. In other words, for all $x\in[0,1]$
we have
\begin{equation}
(-1)^{n}\phi_{n}(x)\geq0
\, . \label{posp}
\end{equation}
For integer $r\geq1$, we further define
\begin{equation}
\varphi_{n}^{(r)}(x)=(-1)^{n}\sum_{s=2}^{n}
\frac{c_{n}^{(s)}x^{s}}{(s-1)^{r}}
\ , \label{vpdef}
\end{equation}
so that the left-hand side of (\ref{posp}) coincides with
$\varphi_{n}^{(1)}(x)$. For $x\in[0,1]$, one has
\begin{equation}
\varphi_{n}^{(r)}(x)\geq0
\, , \label{posp1}
\end{equation}
This conclusion is obtained by induction due to (\ref{posp}) and
\begin{equation}
\varphi_{n}^{(r+1)}(x)=x\int_{0}^{x}\varphi_{n}^{(r)}(z)\,\frac{\xdif{z}}{z^{2}}
\ . \nonumber
\end{equation}
Assuming $\alpha-1=\gamma\in[0,1)$, we now write the power expansion
\begin{equation}
\frac{(-1)^{n}}{\tau_{n\alpha}}=2n^{2}+(-1)^{n}\gamma\,\sum_{s=2}^{n}
\frac{c_{n}^{(s)}}{s-1-\gamma}=
2n^{2}+(-1)^{n}\gamma\,\sum_{s=2}^{n}
\frac{c_{n}^{(s)}}{s-1}\left(1-\frac{\gamma}{s-1}\right)^{\!-1}=
2n^{2}+\sum_{r=1}^{\infty}\varphi_{n}^{(r)}(1)\,\gamma^{r}
\, . \label{intau}
\end{equation}
Due to (\ref{posp1}), the right-hand side of (\ref{intau}) is
positive. That is, we proved $(-1)^{n}\,\tau_{n\alpha}\geq0$ for
$\alpha\in[1,2)$.

\end{document}